\documentclass[prc,aps,nofootinbib,showkeys,showpacs,twocolumn]{revtex4}   
\usepackage{epsfig}  
\usepackage{graphicx}  
\usepackage{amssymb}
\usepackage{amsmath}
\usepackage{color}  
\begin{document}  
  
\title{Systematic of isovector and isoscalar giant quadrupole resonances in normal and superfluid 
deformed nuclei}  
  
\author{Guillaume Scamps}  
 \email{scamps@ganil.fr}  
\affiliation{GANIL, CEA/DSM and CNRS/IN2P3, Bo\^ite Postale 55027, 14076 Caen Cedex, France}  
\author{Denis Lacroix} \email{lacroix@ipno.in2p3.fr}  
\affiliation{Institut de Physique Nucl\'eaire, IN2P3-CNRS, Universit\'e Paris-Sud, F-91406 Orsay Cedex, France}    
    
\begin{abstract}  
The systematic study of isoscalar (IS) and isovector (IV) giant quadrupole responses (GQR) in normal and superfluid 
nuclei presented in [G. Scamps and D. Lacroix, Phys. Rev. {\bf C 88}, 044310 (2013)] 
is extended to the case of axially deformed and triaxial nuclei.  The static and dynamical energy 
density functional based on Skyrme effective interaction are used to study static properties and dynamical response functions
over the whole nuclear chart. Among the 749 nuclei that are considered, 301 and 65
are respectively found to be prolate and oblate while 54 do not present any symmetry axis. For these nuclei,
the IS- and IV-GQR response functions are systematically obtained.  In these nuclei, different aspects related to the interplay between 
deformation and collective motion are studied. We show that some aspects like the fragmentation of the response induced by 
deformation effects in axially symmetric and triaxial nuclei 
can be rather well understood using simple arguments. Besides this simplicity, more complex effects 
show up like the appearance of non-trivial deformation effects on the collective motion damping or the influence of hexadecapole or 
higher-orders effects. A specific study is made on the triaxial nuclei where the absence of symmetry axis adds further complexity to the nuclear response. 
The relative importance of geometric deformation effects and coupling to other vibrational modes are discussed.   
\end{abstract}

 
\maketitle  

\section{Introduction}

Mean-field theories based on the density functional approach \cite{Ben03} have reached in the last decade 
a certain maturity that allows for systematic investigations of nuclear structure effects \cite{Ben06,Ter06,Ber07,Ber09,Rob11}.
Such approaches also permit to address the complexity of phenomena appearing in small and large amplitude collective 
motion from a microscopic point of view. Recently, important progresses have been made to promote
 transport theories based 
on mean-field at the same level as state of the art nuclear structure models.  Time-dependent mean-field approaches are now treating 
the nuclear many-body problem including all terms in the energy density functional theory (EDF)
in a fully unrestricted three dimensional space \cite{Sim10,Sim12,Mar13,Sek13,Ste04}. More recently, important efforts have been made to include 
pairing correlations \cite{Has07,Ave08,Ste11,Eba10}. Application of the time-dependent EDF (TD-EDF) theories usually focus on rather specific problems 
related to selected phenomena in a small set of nuclei. We believe however, that such approach can now, similarly 
to nuclear structure models, be applied to perform systematic studies.  
The aim of the present work, following Ref. \cite{Sca13a}, is to make systematic applications 
of TD-EDF based on Skyrme functional in order to make global assessments on their applicability, predictive power and
eventually give deeper physical insight in specific phenomena. 

Collective motion in nuclei, being at the crossroad between nuclear structure  and nuclear reactions, 
are a perfect laboratory. Important progresses have been made recently to extend quasi-particle 
random phase approximation (QRPA) to the treatment of nuclei that are deformed in their ground states 
\cite{Yos06,Hag04,Per08,Pen09,Per11,Paa03,Ter04,Los10,Yos13,Nik13,Lia13}. Most recent QRPA applications are able to treat 
axially deformed systems over the whole nuclear chart.  The description of systems without any symmetry axis, i.e. triaxial nuclei, 
remains a challenge in the field. From that point of view, truly time-dependent approaches like TDHFB or TDHF+BCS appear as
promising methods.  Here, we consider the example of the  isoscalar and isovector giant quadrupole resonances. 

\section{GQR in deformed nuclei: generalities}

\begin{figure}[!ht]   
	\centering\includegraphics[width=1.0\linewidth]{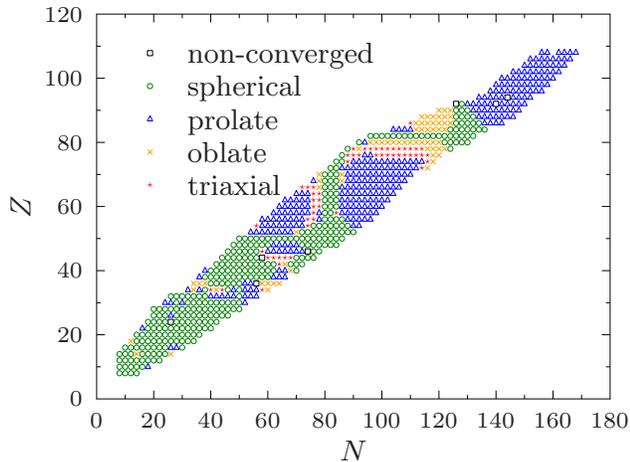}  
	\caption{(Color online) Nuclei retained in the present study. The different markers 
	refer to the deformation type found in the ground state using the {\rm EV8} code with SkM*
	functional. The black open squares indicate nuclei for which an insufficient convergence has been obtained in {\rm EV8} (see text).} 
	\label{fig:map_def} 
\end{figure} 
\begin{figure}[htbp] 
	\centering\includegraphics[width=0.9\linewidth]{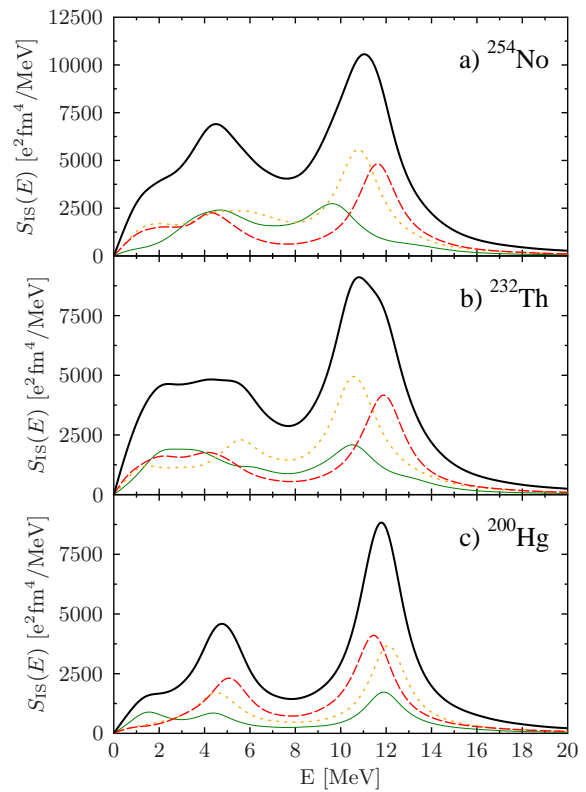}  
	\caption{ (Color online)  Illustration of the energy splitting of the different $|K|$ components in the 
	IS-GQR strength function due to deformation. The cases of a) $^{254}$No , b) $^{232}$Th and c) $^{200}$Hg are considered. 
	In each case, the $|K|=0$ (green open squares line), 
	 $|K|=1$ (orange short dashed line) and $|K|=2$ (red long dashed line) are shown as well as the total strength 
	(black solide line) obtained by summing up the different contributions. Note that the two first nuclei are prolate 
	while the last one is oblate. A smoothing parameter $\gamma=1$ MeV is used in this figure. }
		\label{fig:fewA} 
\end{figure}

\begin{figure}[htbp] 
	\centering\includegraphics[width=\linewidth]{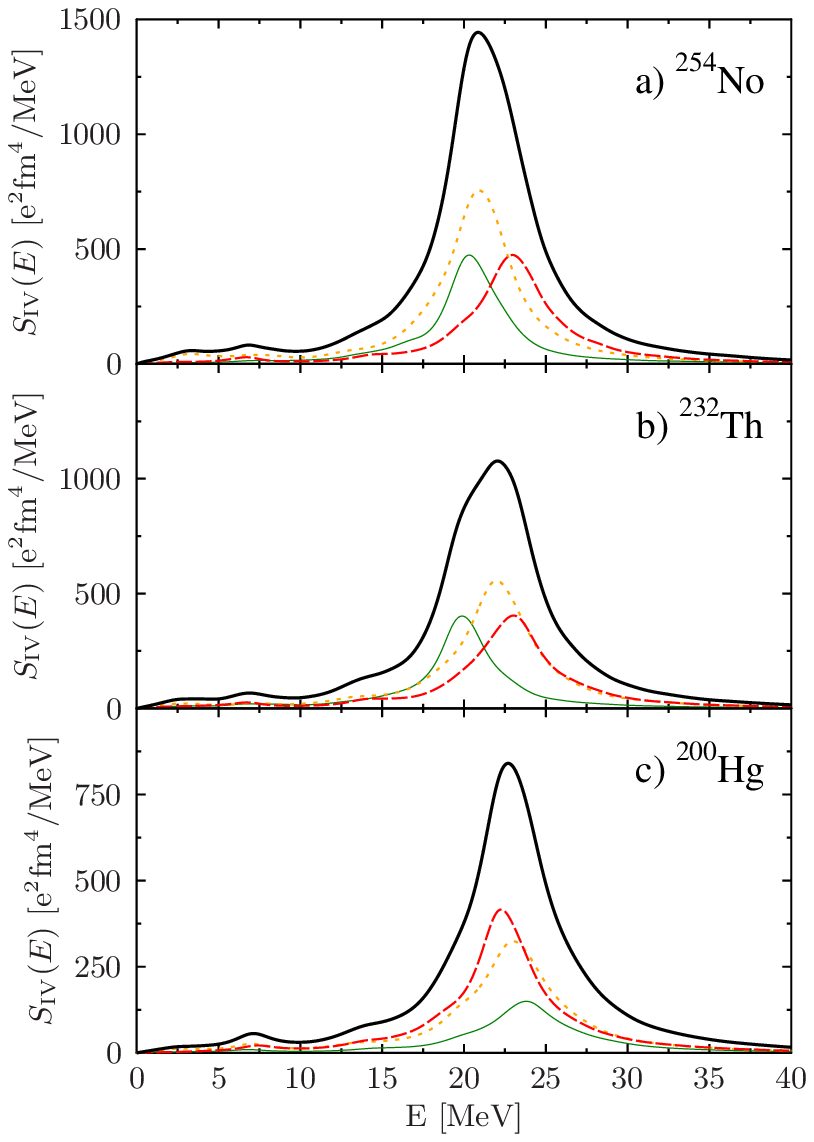}  
	\caption{ (Color online) Same as Fig. \ref{fig:fewA} for the IV-GQR. } 
	\label{fig:fewA-IV} 
\end{figure}
In the present article, the isoscalar and isovector GQR responses of 749 nuclei have been 
obtained using the static and dynamical Skyrme EDF with the SkM* interaction in the 
mean-field channel. The {\rm EV8} code \cite{Bon05} is used to initialize the different nuclei. 
All technical details associated to the initial conditions can be found 
in Ref. \cite{Sca13a,Sca13b}.  Due to the symmetries imposed in  {\rm EV8}, only deformations 
associated to the parameters 
$\beta_\lambda$ with even multipolarities can be considered, i.e. $\lambda= 2$, $4$, $6\dots$ 
Nevertheless, this code is sufficiently 
general to lead to spherical, prolate, oblate or triaxial shapes.  A summary of the different 
shapes considered in the present work is given in Fig. \ref{fig:map_def}. 
\begin{figure}[htbp] 
	\centering\includegraphics[width=\linewidth]{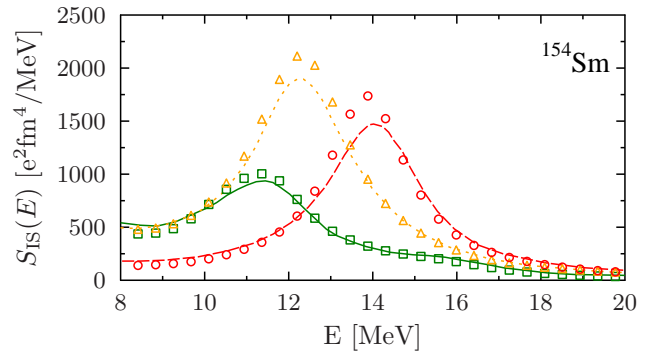}  
	\caption{ (Color online)  Comparison between the TDHF+BCS IS-GQR strength distributions obtained 
	in $^{154}$Sm for $|K|=0$ (green open squares line), 
	 $|K|=1$ (orange short dashed line) and $|K|=2$ (red long dashed line).  The different curves are the 
	 corresponding QRPA results from Ref. \cite{Yos13}. Note that the TDHF+BCS strengths have been 
	 smoothed with a width of $2$ MeV consistently with the QRPA results.} 
	\label{fig:sm154} 
\end{figure}

The nuclear response is studied by applying quadrupole boost operators, denoted by $\hat Q^{IS/IV}_{2K}$ 
to the ground state wave-function, where $K = 0,$ $\pm 1$, $\pm 2$ (see Eqs. (7-14) of Ref. \cite{Yos13}). 
Note that in practice, we used linear combinations of these operators that have the advantage to be real (see appendix \ref{app:excitation}).  Then, the dynamical evolution is followed 
in time using the recently developed TDHF+BCS approach \cite{Sca13}. A preliminary study 
of the IS-GQR and IV-GQR focusing on spherical nuclei was done in Ref. \cite{Sca13a} where details of the dynamical evolution and on the methodology to extract the response function 
can be found. Here, we concentrate on deformed nuclei. For these nuclei, the response 
function associated to each $K$ are expected to differ from each other. More specifically, if the 
nucleus has an axial symmetry (prolate and oblate nuclei), the $K=+1$ and $K=-1$ (resp. $K=+2$ and $K=-2$) will be
degenerated 
while a splitting of the $|K|=0$, $|K|=1$ and $|K|=2$ will occur depending  on the deformation type \cite{Suz77,Abg80,Jan83,Nis85}.   
This effect illustrated for few nuclei in Figs. \ref{fig:fewA} (IS-GQR) and \ref{fig:fewA-IV} (IV-GQR) 
will be analyzed in detail below. 
For nuclei with triaxial deformation all response functions are expected to differ from each others. Note that the quality of the calculation have been systematically tested by comparing the 
sum rule obtained with by integrating the strength function and the analytical expression given in appendix \ref{sec:sumrule}.
In all considered nuclei a maximum of 0.5 \% of error has been found that could be assigned mainly to the finite mesh step (see discussion in Ref. \cite{Sca13a}).
It is worth mentioning that negative tails in the strength function conjointly to a breakdown of the EWSR is served if 
a perfect convergence of the self consistent Hartree-Fock initial equation is not achieved with a very good precision. 
For 7 nuclei, reported by black open squares in Fig. \ref{fig:map_def}, 
it was impossible to achieve a properly converged static solution. These nuclei are removed from the present analysis.

It should be mentioned that the $|K|=0$, and $|K|=2$ excitation are vibrational excitation while the $|K|=1$ excitation excite 
also rotations. This is clearly seen in the time-dependent evolution corresponding to this excitation
that leads to a slow rotation of the nucleus (see appendix \ref{sec:rotation}) leading to a low frequency component 
in the response function. It should be noted that this contribution is spurious. Indeed, the initial state and any rotated 
state have the same mean-field energy.  Therefore, in a configuration mixing picture, the true many-body state should be 
obtained by mixing all possible orientations of the nucleus which is equivalent to restoring the total 
angular momentum.  To remove the spurious contribution, the spectra is first corrected from the rotation prior 
to the signal Fourier transformation. An illustration of the technique is given in appendix   \ref{sec:rotation}.

\subsection{Comparison with QRPA}
\begin{figure}[htbp] 
	\centering\includegraphics[width=0.7\linewidth]{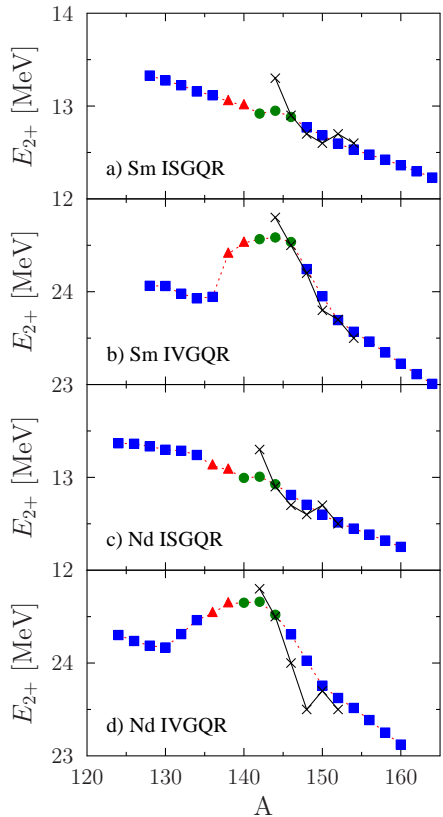}  
	\caption{ (Color online) Comparison between the mean GQR energy obtained by fitting the 
	strength distribution with a Lorentzian shape around the collective energy. 
	Mean energies of the IVGQR and ISGQR for the Sm isotopic chains are presented in panel 
	(a) and (b) respectively. The   IVGQR and ISGQR for the Nd isotopic chains are presented in panel 
 (c)  and (d).
	The fit has been performed on the strength distribution smoothed by a Lorentzian 
	distribution of 2 MeV width. The crosses correspond to the results given in Ref. \cite{Yos13}. 
	The different symbols correspond to the spherical symmetric (green filled circles), prolate (blue filled squares)
	and triaxial (red filled triangles) nuclei.  } 
	\label{fig:comp_e} 
\end{figure}
The response of superfluid nuclei is generally studied using the QRPA 
approach that is the small amplitude limit of TDHFB. In the present work, we do not 
perform the full TDHFB evolution but approximate the pairing field to obtain 
a simpler TDHF+BCS version.  This approximation was found to work remarkably 
well in the case of spherical nuclei. Here we show that this conclusion still holds 
in the case of deformed systems.  For comparison, we took one of the latest  
results obtained with state of the art QRPA approach allowing for 
axial deformation using Skyrme functional  \cite{Yos13}. Note that, the same SkM* functional 
has been used also in the QRPA case but the pairing interaction used here is the 
surface interaction of Ref. \cite{Ber06} that differs from Ref. \cite{Yos13}.

In Fig.  \ref{fig:sm154}, the response function shown in Ref.   \cite{Yos13} is compared with our TDHF 
results for the different $K$ values.  A very good agreement with the QRPA result is obtained 
even if the BCS approximation is made in our calculation. In particular the peak position are perfectly 
reproduced. 
This confirm our previous finding 
\cite{Sca13a} that, except in the low-lying energy sector, the TDHF+BCS approximation provides a powerful 
theory to account for pairing effects in giant resonances. It should be noted that a small physical damping of the 
order of $0.5$ MeV is present in the QRPA results that we do not have in TDHF+BCS. Indeed, the two 
calculations perfectly match with each others if a Lorentzian smoothing width $\gamma=2.5$ MeV is used  
instead of $\gamma=2.0$ MeV. This seems to be systematically the case for the IS-GQR but not for the IV-GQR 
where a perfect agreement is found without increasing artificially the smoothing parameter.
\begin{figure}[!ht] 
	\centering\includegraphics[width=0.7\linewidth]{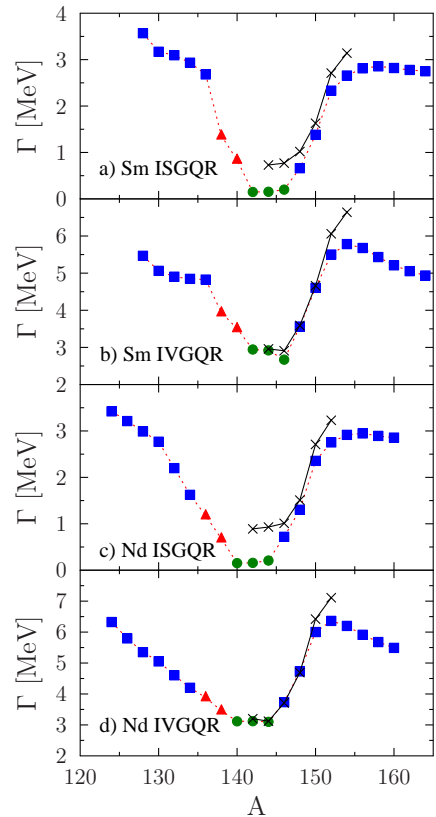}  
	\caption{(Color online)  Widths of the IS and IV-GQR resonances in the Sm and Nd isotopic chains. The convention are the same as
	in Fig. \ref{fig:comp_e}. 
	} 
	\label{fig:comp_gamma} 
\end{figure}

In Fig. \ref{fig:comp_e} and \ref{fig:comp_gamma},  we systematically compare the 
mean energy, denoted by $E_{2^+}$  
and total width $\Gamma$ of the total IS-GQR and IV-GQR response for the Nd and Sm 
isotopic chain, the total strength being obtained by summing all $K$ 
response function. The values of $E_{2^+}$ and $\Gamma$ have been obtained here by fitting the 
main collective peak with a single Lorentzian distribution.  Again, for both IS and IV excitations a rather 
good agreement for the energy and width is found with the QRPA results. Note that a more detailed discussion 
on the width evolution is made below. 

In all the comparisons we have made with QRPA, very good 
agreements were found. 
In our opinion, this agreement is due to the fact that the main effect of pairing on normal
giant resonances stems from its influence to decide the ground state initial deformation and initial fragmentation of single-particle occupation around the Fermi energy. These   
effects are rather well described in the BCS approach for not too exotic nuclei. True dynamical pairing effects 
are only affecting the response function in the low-lying sector or the response that explicitly 
involve the anomalous density and change particle number, like pairing vibrations.   

\section{Axially symmetric nuclei} 

\begin{figure}[!ht] 
	\centering\includegraphics[width=1.0\linewidth]{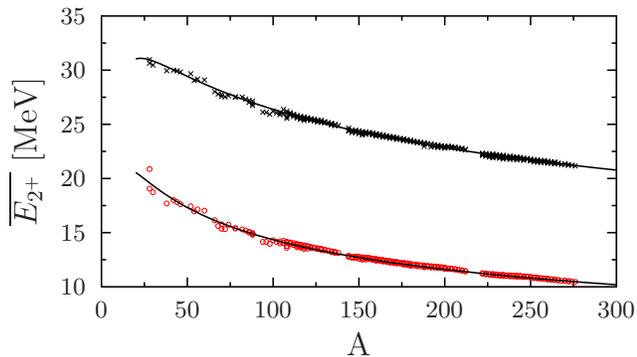}  
	\caption{ (Color online)  Mean value of the principal peak as a function of the mass $A$ for the IS excitation (red open circles ) and the IV excitation (black crosses). The two lines correspond to the formula given by Eq. (17) 
	of Ref. \cite{Sca13a} fitted on spherical nuclei. } 
	\label{fig:Mean_E_fctA} 
\end{figure}

In the present section, we concentrate on nuclei that present an axis of symmetry. With the SkM*, this corresponds 
to 366 nuclei (301 prolate and 65 oblate) over the nuclear chart (see Fig. \ref{fig:map_def}). Due to the symmetry, only three 
spectra need to be considered, one for each $|K|$ value. Following, Ref. \cite{Sca13a}, for each nuclei and $|K|$ values,  the mean collective 
energy and width can be obtained using a fitting procedure with a single Lorentzian function of the corresponding strength. 
Before regarding each spectra separately, let us mention that the average energy obtained with:
\begin{align}
\overline{ E_{2^+}}&=\frac{1}{5} \Big(E[d_{z^2}^{K=0}] + E[d_{xy}^{|K|=2}] + E[d_{xz}^{|K|=1}] \nonumber \\
&+ E[d_{yz}^{|K|=1}] + E[d_{x^2-y^2}^{|K|=2}] \Big) \nonumber \\
&=\frac{1}{5} \left( E_{|K|=0} + 2 E_{|K|=1} + 2 E_{|K|=2}\right) ,
\end{align}
where $E[F]$ is a generic expression for the mean energy associated with the excitation operator $F$ (see appendix \ref{app:excitation} for the expressions of different operators) and 
where the last expression only applies to the case of axially symmetric nuclei. The average quantity $\overline{ E_{2^+}}$
is shown as a function of the mass of the nuclei for the IS- and IV-GQR in Fig. \ref{fig:Mean_E_fctA} and are compared 
with the fitted expression for spherical nuclei (Eqs. (17) of Ref. \cite{Sca13a}).
We see that the average property in deformed nuclei perfectly match the trends obtained in spherical nuclei.

\subsection{Splitting of collective energies}

The splitting of collective energies depending on the $|K|$ value due to 
 deformation is a well established property \cite{Boh97,Har01} and is usually 
seen in QRPA calculation \cite{Yos13,Per08,Hag04,Los10}.
Several simple models have been proposed to qualitatively understand this splitting. Among them, one 
could quote the adiabatic cranking approximation \cite{Abg80}, the scaling and fluidynamical approach \cite{Nis85}
and the variational approach of Ref. \cite{Jan83}.  One expect generally two effects that influence the splitting. 
The first one is purely geometric and will be directly connected to the radius in the direction along 
which the excitation is applied. This effect is the dominant one at small quadrupole deformation. 
At larger deformation, we do expect to see the effect of the coupling between GQR and GMR modes. 
 First indication that this coupling explicitly shows up in QRPA can be found in Ref. \cite{Yos13}.
Below, a larger scale analysis of these aspects is made.    

\begin{figure}[!ht] 
	\centering\includegraphics[width=1.0\linewidth]{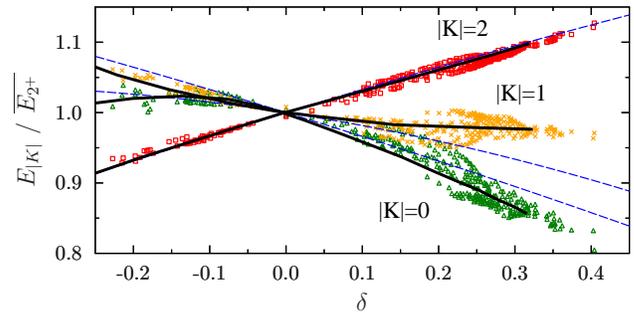}  
	\caption{ (Color online)  Main peak energy of the IS-GQR as a function of $\delta$ for $|K| = 0$ (green open triangle), 
	$|K| = 1$ (orange crosses) and $|K| = 2$ (red open squares).  The fluid-dynamical model (directly taken from Fig. 1-a of Ref. 
	\cite{Nis85}) are shown with thick black solid lines while the simple scaling model is shown by dashed blue lines.} 
	\label{fig:comp_Nishizaki}
\end{figure}

In Fig. \ref{fig:comp_Nishizaki}, the energies of the $|K| =0$, 1 and 2 main peaks are displayed as a function of the deformation parameter: 
\begin{align}
\delta &= \frac{3\sqrt{ \langle Q_{20}\rangle^2 + 3 \langle Q_{22}\rangle^2}}{4 A \langle r^2\rangle},  \label{eq:delta}
\end{align}
with
\begin{align}
\langle r^2 \rangle &= \frac1A \langle \Psi | \sum_{i=1}^A {\hat r_i}^2 | \Psi \rangle,  \\
\langle Q_{20} \rangle &=  \langle \Psi | \sum_{i=1}^A (2 {\hat z_i}^2 -{\hat y_i}^2 - {\hat x_i}^2) | \Psi \rangle,  \\
\langle Q_{22} \rangle &=  \langle \Psi | \sum_{i=1}^A ({\hat x_i}^2 - {\hat y_i}^2) | \Psi \rangle. 
\end{align}
In this figure, we do observed the expected hierarchy in the energies, i.e. for the prolate nuclei ($\delta > 0$) the 
collective $|K| = 2$ energy is above the $|K| = 0$ while the $|K| = 1$ lies in-between. For the oblate cases  ($\delta < 0$), 
the hierarchy is inverted.
The deformation splitting can be compared with the scaling model, shown by dashed lines in Fig. \ref{fig:comp_Nishizaki},
where an analytical $\delta$-dependence of the splitting is given by \cite{Nis85}:   
\begin{align}
\frac{E_{|K|=0}}{\overline{E_{2+}}} &= 1 - \frac13 \delta - \frac1{18} \delta^2 , \label{eq:0delta}\\
\frac{E_{|K|=1}}{\overline{E_{2+}}} &= 1 - \frac16 \delta - \frac{13}{72} \delta^2 , \label{eq:1delta}\\
\frac{E_{|K|=2}}{\overline{E_{2+}}} &= 1 + \frac13 \delta - \frac{1}{18} \delta^2 .\label{eq:2delta}
\end{align}
This expressions give properly the average trend either for the $|K|=2$ case over the entire range of $\delta$ and 
for the other $|K|$ when the deformation is very small.
The fluid-dynamical model (thick solid line in Fig. \ref{fig:comp_Nishizaki}) reproduce very well the average trend. In this model the coupling with the giant monopole mode is accounted for and the spurious rotational mode induces the difference between the scaling and hydrodynamical model in the $|K|=1$ channel. 
The excellent agreement with the average TDHF+BCS splitting shows the importance of coupling effects for the $K=0$ mode and the influence of the rotation for the $|K|=1$ collective energy.

\begin{figure}[!ht] 
	\centering\includegraphics[width=1.0\linewidth]{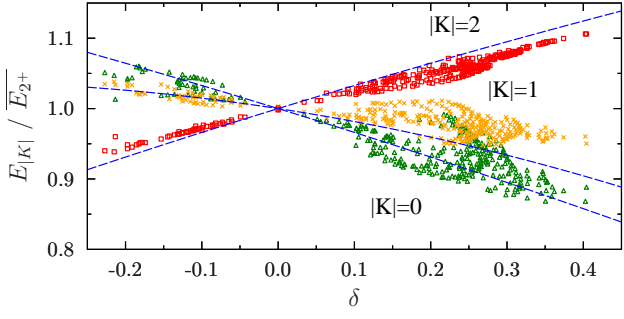}  
	\caption{ (Color online) Same as figure \ref{fig:comp_Nishizaki} for the IV-GQR. 
	Note that we also show the scaling model result of the IS-GQR as a reference (blue dashed line). } 
	\label{fig:comp_Nishizaki_IVGQR} 
\end{figure}

In Fig. \ref{fig:comp_Nishizaki_IVGQR}, for completeness, 
we show the energy splitting observed for the IV-GQR that is compared also 
with the same scaling model. Note that, for that case, we do not have the fluid-dynamical model result. We see that the 
hierarchy in energy is respected.

Independently of the underlying physics, the average trend of the energy for each $|K|$ can be directly 
obtained by fitting the  evolution of the energies displayed in Figs. \ref{fig:comp_Nishizaki} (IS-GQR) and \ref{fig:comp_Nishizaki_IVGQR} (IV-GQR)
with a second order polynomial in $\delta$:
\begin{align}
E^{IS/IV}_{|K|}&=  \overline{E^{IS/IV}_{2+} }\left( 1 + a^{|K|}_1 \delta + a^{|K|}_2 \delta^2 \right). \label{eq:fit2}
\end{align}
Results of the fits are given in table \ref{tab:fit}. 
\begin{table}
\begin{tabular}{|c||c|c|}
\hline
& $~~~a^{|K|}_{1}~~~$ & $~~~a^{|K|}_{2}~~~$   \\
 \hline \hline
IS K=0 & -0.240 & -0.672 \\
IS K=1 & -0.167 & 0.308 \\
IS K=2 & 0.305 & -0.036 \\ 
 \hline
\end{tabular}
\hspace*{0.cm}
\begin{tabular}{|c||c|c|}
\hline
& $~~~a^{|K|}_{1}~~~$ & $~~~a^{|K|}_{2}~~~$   \\
 \hline
 \hline
IV K=0 & -0.277 & -0.052 \\
IV K=1 & -0.111 & 0.032 \\
IV K=2 & 0.250 & -0.006 \\
 \hline
\end{tabular}

  \caption{Values of the parameters $ a^{|K|}_1$ and $ a^{|K|}_2$  by fitting the energies displayed in Figs. 
  \ref{fig:comp_Nishizaki} or \ref{fig:comp_Nishizaki_IVGQR} with a second order polynomial in $\delta$ given by 
 Eq.  (\ref{eq:fit2}).}
\label{tab:fit}
\end{table}

\begin{figure}[!ht] 
	\centering\includegraphics[width=0.8\linewidth]{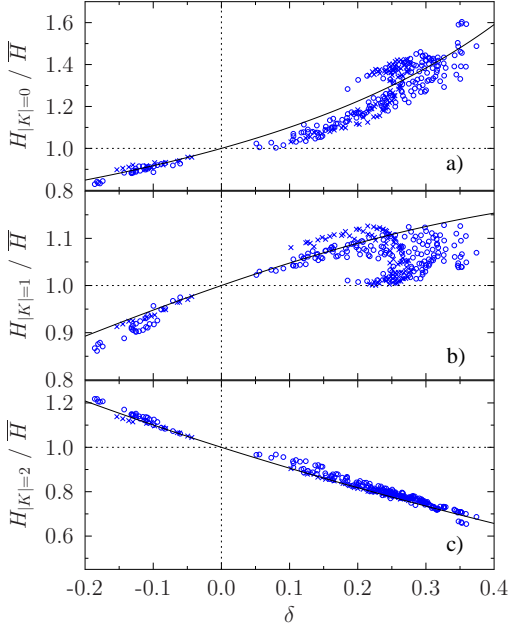}  
	\caption{ (Color online)  Heights of the main collective IS-GQR peak for the components a) $|K| =0$ , b) $|K|=1$ (middle) and  c) $|K|=2$. The solid lines are the average trends obtained using Eqs. (\ref{eq:peakdelta}) togethers with the fitted mean collective energies (Eq. (\ref{eq:fit2})). } 
	\label{fig:heightIS} 
\end{figure}
\begin{figure}[!ht] 
	\centering\includegraphics[width=0.8\linewidth]{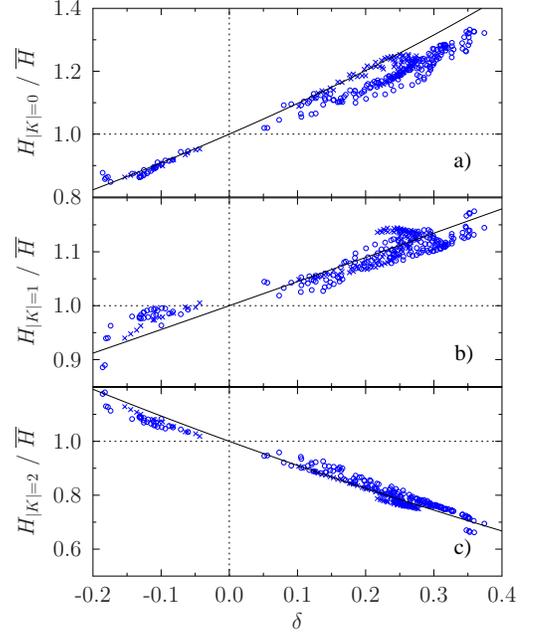}  
	\caption{ (Color online) Same as Fig. \ref{fig:heightIS} for the IV-GQR. }  
	\label{fig:heightIV} 
\end{figure}

\subsection{Effect of deformation on the collective peaks height}

Deformation not only affects the peaks position but also their relative importance in the total
strength function. This is clearly seen in Fig. \ref{fig:heightIS} where the peaks height of the main collective peak
for each $|K|$, denoted by $H_{|K|}$ and normalized to the average quantity: 
 \begin{align}
{\overline H} = \frac{1}{5} \left(H_{|K| = 0} + 2 H_{|K| = 1} + 2  H_{|K| = 2} \right),
\end{align}
are shown as a function of the deformation parameters. We clearly see 
that the $|K|=2$ contribution is reduced in favor of the $|K|=0$ and to a lesser extend in factor of $|K|=1$. 
In the prolate nuclei, the opposite is seen. In Fig. \ref{fig:heightIV}, the same trends are seen for the IV-GQR 
case.  

These trends can be rather easily understood using sum-rule arguments. Considering the IS-GQR case,
starting from the sum-rules derived  in appendix \ref{sec:sumrule} and using the definition of $\delta$, we deduce that the $m_1$
sum-rules verifies 
\begin{align}
m^{|K|=0}_1 &= \overline{m_1}  ~(1+2/3\delta),  \\
m^{|K|=1}_1 &= \overline{m_1}  ~(1+1/3\delta), \\
m^{|K|=2}_1 &= \overline{m_1}  ~(1- 2/3\delta), 
\end{align}
where $ \overline{m_1}  = 5e^2 \hbar^2 A \langle r^2 \rangle / (4 \pi m)$. Making the assumption 
that the collective response has a single collective peaks with no fragmentation 
leads to the relationship:
\begin{align}
E_{|K|} H_{|K| } &\simeq m^{|K|}, ~~~~{|K|=0,~1,~2}.  
\end{align}
If in addition we simply assumes that $\overline{ E_{2^+}} ~{\overline H} \simeq \overline{m_1} $,
the peak heights  approximately verify:
\begin{align}
\left\{
\begin{array}{c} 
  H_{|K|=0}/{ \bar H}= (1+2/3\delta)/E_{|K|=0}(\delta) \\
  \\
  H_{|K|=1}/{ \bar H}= (1+1/3\delta)/E_{|K|=1}(\delta) \\
  \\
  H_{|K|=2}/{ \bar H}= (1-2/3\delta)/E_{|K|=2}(\delta) 
 \end{array}
 \right. .
 \label{eq:peakdelta}
\end{align}
To check the validity of these formulas, we have reported in Fig. 
\ref{fig:heightIS} with solid lines, the resulting dependences 
of the peak energy with $\delta$ where $E_{|K|}$ is replaced by the quantity (\ref{eq:fit2}).
The solid lines are able to globally describe the peaks height increase 
or decrease. Similar formulas can also be derived for the IV-GQR and the resulting trends are also shown 
by solid lines in Fig. \ref{fig:heightIV}.

\subsection{Effect of deformation on fragmentation and damping} 

\begin{figure}[!ht] 
	\centering\includegraphics[width=\linewidth]{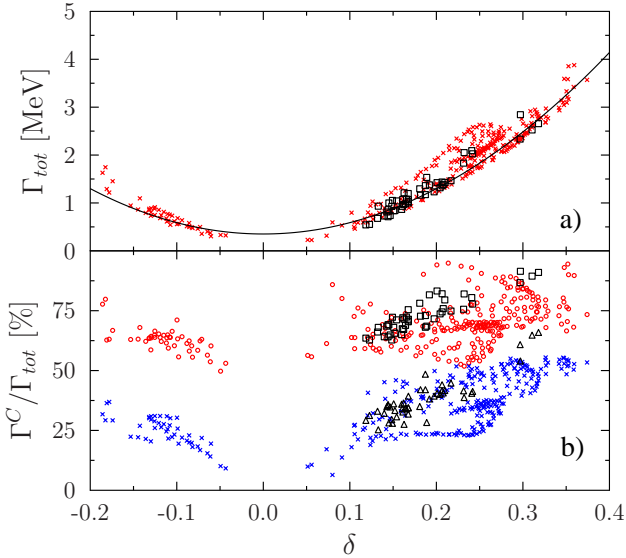}  
	\caption{(Color online) 
	Top: Total damping width $\Gamma_{\rm tot}$ of the IS-GQR is plotted a function of the deformation parameter $\delta$ in 
	axially symmetric nuclei (red crosses). 
	The black solid line is the result of the fit (Eq. (\ref{eq:fitgamtot})).  The width for nuclei with triaxial deformation are also reported with black open squares. 
	Bottom: 	The quantity $\Gamma^C_{\rm tot}/\Gamma_{\rm tot}$,  
	computed from formula (\ref{eq:gg}) is plotted as a function of 
        $\delta$ (red open circles). For comparison, the quantity $\Gamma^C_{E}/\Gamma_{\rm tot}$ that only 
        account for the energy splitting of the different $|K|$ components is also shown (red crosses). The corresponding 
        quantities for triaxial 
        nuclei are displayed by  black open squares and black open triangles.}
	\label{fig:widthtot} 
\end{figure}
\begin{figure}[!ht] 
	\centering\includegraphics[width=0.8\linewidth]{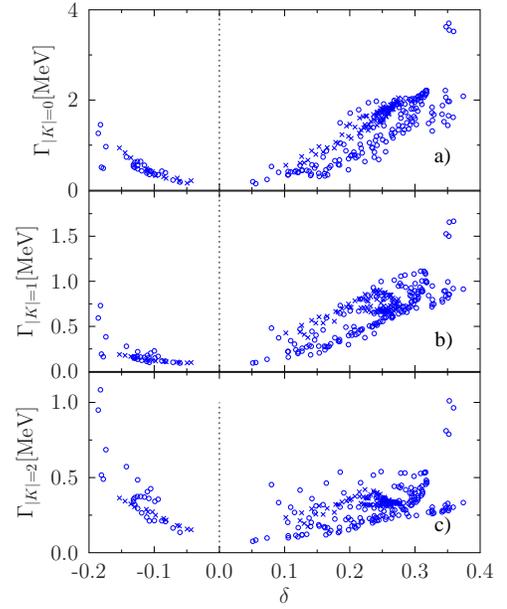}  
	\caption{ (Color online) Widths of the main collective IS-GQR peak for the components a) $|K|=0$ , b) $|K|=1$ and  c) $|K|=2$. Note that the smoothing parameter has been removed from the width as in Ref. \cite{Sca13a}
	to focus on the physical width. }
	\label{fig:widthIS} 
\end{figure}

In the IS-GQR channel, the physical width obtained with TDHF+BCS has been shown in Ref. \cite{Sca13a} 
to be zero or almost zero in medium and heavy-mass spherical nuclei. In deformed systems, due to the splitting of the collective energy
illustrated in Fig. \ref{fig:comp_Nishizaki}, the total strength function is fragmented and will systematically acquire a width. 
In Fig. \ref{fig:widthtot}  
the total width $\Gamma_{\rm tot}$, obtained by a fit of the total strength with a single Lorenzian, is systematically reported 
as a function of the deformation parameter $\delta$ for nuclei with $A>100$. Note that,  in 
Fig. \ref{fig:widthtot} as well as in all widths shown below, 
the width corresponds to the physical width where the contribution of the smoothing procedure 
has been removed \cite{Sca13a}. As expected, the width increases as a function of $\delta$. The $\delta$-dependence 
turns out to be compatible with a simple quadratic dependence :
\begin{align}
\Gamma_{\rm tot} &= 0.35 + 23.7 \delta^2~~[{\rm MeV}] ,  \label{eq:fitgamtot}
\end{align}  
shown in top panel of Fig. \ref{fig:widthtot} by solid line.

A detailed analysis shows that the total width cannot be understood by invoking the energy splitting of the 
different $|K|$ components only. In particular each channel $|K|$ acquires a physical width that is reported in Fig. \ref{fig:widthIS}
as a function of the deformation parameter.



Again, in this figures, the width are obtained by a fitting procedure for each $|K|$ value.   We see that there is a hierarchy in the evolution of the width $\Gamma_{|K|=0} > \Gamma_{|K|=1} > \Gamma_{|K|=2}$. In particular, there is a strong increase in the $|K|=0$  case. 
 
To study the different contributions acting on the giant resonance damping, we can use the approximate formula derived in appendix \ref{sec:width}:
\begin{align}
\Gamma^2_{\rm tot} &= \sum_K \frac{H_{|K|}}{\overline{H}}  \Gamma^2_{|K|} +  \Gamma^2_E , \label{eq:gg}
\end{align}
where $ \Gamma^2_E$ corresponds to the contribution of the energy splitting to the total width. It can be expressed 
as a functional of the main collective peak energies and heights (see appendix \ref{sec:width}
for more details).  The value of $\Gamma_{\rm tot}$, denoted by $\Gamma^C_{tot}$ and computed from the values of widths, peak heights and collective energies deduced in each channel $|K|$ can be compared to the $\Gamma_{\rm tot}$ obtained 
directly by fitting the total strength that is displayed in Fig. \ref{fig:widthtot}. In bottom panel of this figure, the quantity 
$\Gamma^C_{tot} / \Gamma_{\rm tot}$ is displayed as a function of $\delta$. A perfect agreement between the two ways to 
extract the total width would correspond to $\Gamma^C_{tot} / \Gamma_{\rm tot} \simeq 1$. We see in this figure that 
the calculated value is satisfactory but slightly lower.  Such a discrepancy is not surprising due to the fact that (i) it is 
expected to be valid when a single peak exists in the  total strength which is very unlikely at large deformation (ii) to derive Eq. (\ref{eq:gg}), 
a gaussian shape has been assumed for the collective peak that might induce a systematic error (see appendix \ref{sec:width}). 
Nevertheless, formula  (\ref{eq:gg}) carries interesting informations since it allows to disentangle the different contributions 
to the giant resonance damping. In Fig.  \ref{fig:widthtot} (bottom panel), 
the quantity $\Gamma_{E} / \Gamma_{\rm tot}$ is also reported (blue crosses) showing that half of the damping is due to 
the energy splitting while the rest comes from the physical damping of the IS-GQR that strongly increases as the deformation changes. 

\begin{figure}[!ht] 
	\centering\includegraphics[width=\linewidth]{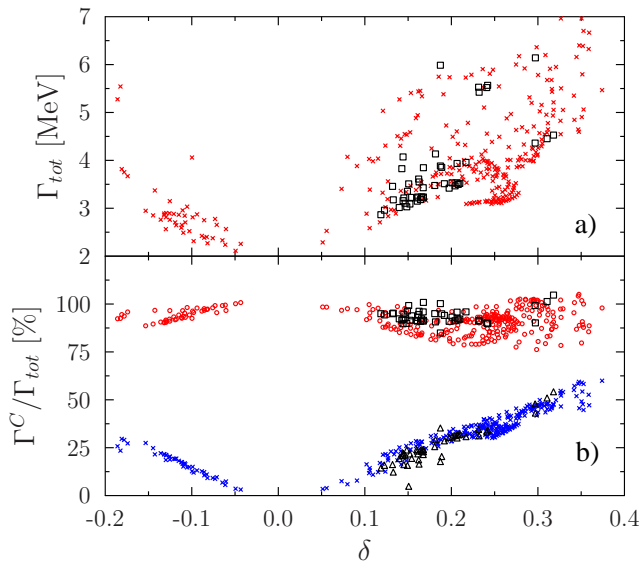}  
	\caption{(Color online) Same as Fig. \ref{fig:widthtot}  for the IV-GQR.}
	\label{fig:widthtotIV} 
\end{figure}

\begin{figure}[!ht] 
	\centering\includegraphics[width=0.8\linewidth]{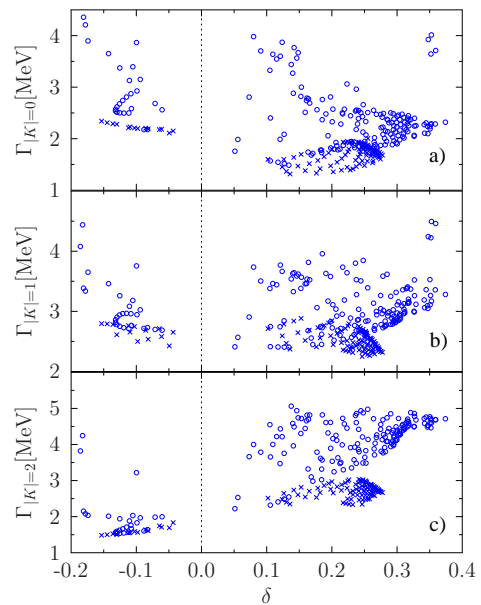}  
	\caption{ (Color online) Same as Fig. \ref{fig:widthIS} for the IV-GQR.}
	\label{fig:widthIV} 
\end{figure}

A similar analysis can be made in the IV-GQR case, the fitted total width, the reconstructed one (Eq. (\ref{eq:gg})) and the splitting in energy contribution are shown in 
Fig. \ref{fig:widthtotIV} 
while the widths in each $|K|$ channels are shown in Fig. \ref{fig:widthIV}. 
The situation in the IV-GQR is slightly different compared to its IS counterpart. First, already 
in the spherical nuclei, the GR possesses an intrinsic physical width. In deformed nuclei, we do see also an increase 
of the damping with the deformation. However much more fluctuations arise compared to the IS case. These fluctuations 
points out (i) that highly non-trivial effects occur in the damping mechanism (ii) that other important effects than the deformation are important in the damping. When Eq. (\ref{eq:gg}) is used to reconstruct the total width from individual $|K|$ contribution,
the approximate expression seems to be in better agreement with the directly fitted $ \Gamma_{\rm tot}$. In that case, less than 50 \% of the width increase can be attributed to the energy splitting. 
 



\subsection{Higher order deformation effect}

\begin{figure}[!ht] 
	\centering\includegraphics[width=1.0\linewidth]{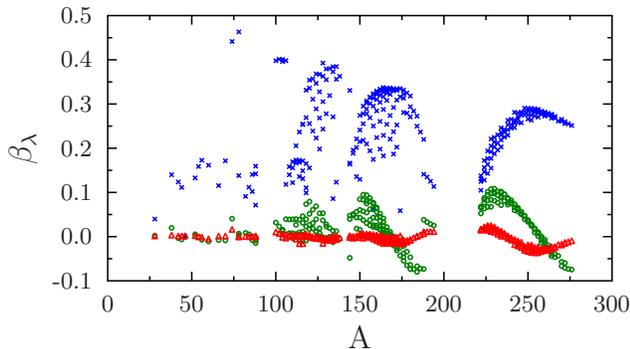}  
	\caption{ (Color online) Deformation parameters as a function of the mass for the SkM* functional. The $\beta_2$ (blue crosses), 
	$\beta_4$ (green circles) and $\beta_6$ (red triangles)  parameters are shown. } 
	\label{fig:beta_246_exp} 
\end{figure}

\begin{figure}[!ht] 
	\centering\includegraphics[width=1.0\linewidth]{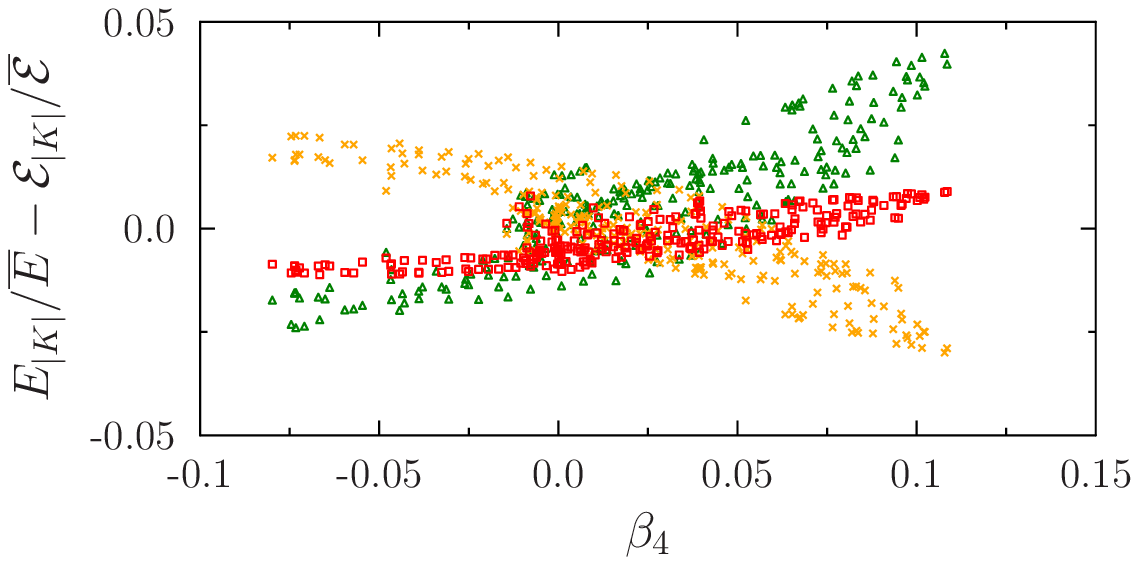}  	
	\centering\includegraphics[width=1.0\linewidth]{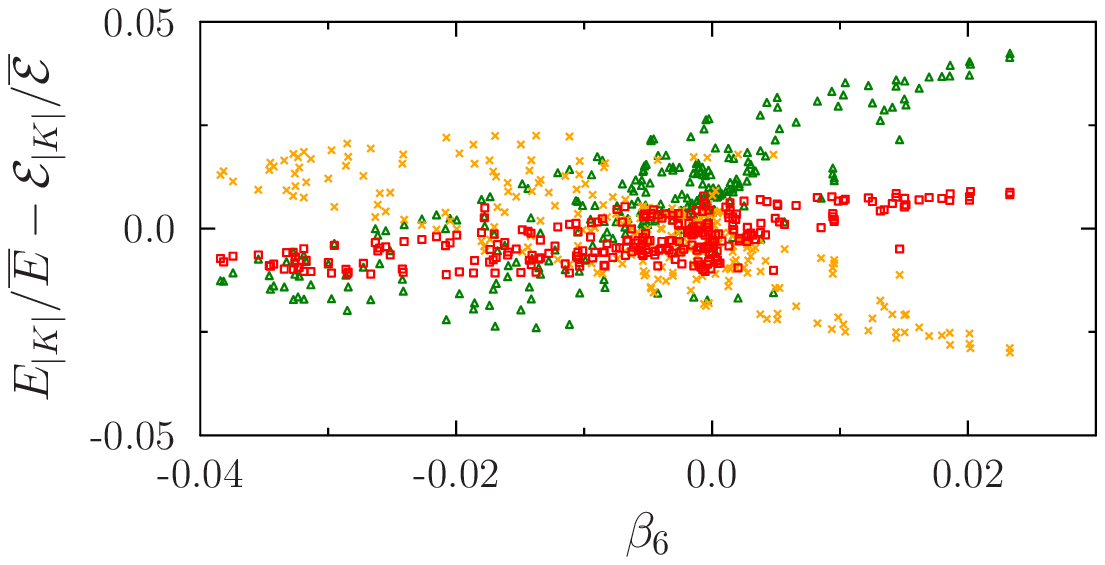}  
	\caption{ (Color online) Deviation of different collective peaks obtained with TDHF+BCS and the reference results, denoted by 
	${\cal E}_{|K|}$, given by the fluid-dynamical model as a function of $\beta_4$ (top) and $\beta_6$ (bottom) 
	for the IS-GQR resonance. The different markers
	correspond to $|K| = 0$ (green open triangle), 
	$|K| = 1$ (orange crosses) and $|K| = 2$ (red open squares).}
	\label{fig:effect_beta_4} 
\end{figure}

From Fig. \ref{fig:comp_Nishizaki}, we see that the fluid-dynamical model provides a rather good description
of the average splitting trends. However, from this figure, we also remark large dispersion around the average trends.   
One origin of this dispersion is the existence of hexadecapole and higher order deformation in the considered 
nuclei (see Fig. \ref{fig:beta_246_exp}). Many nuclei present non-negligible hexadecapole deformation
and also, to a less extent non-zero values of $\beta_6$. In Fig. \ref{fig:effect_beta_4}, a clear correlation is observed 
between the peak energies and the two parameters $\beta_4$ and $\beta_6$ obtained with the method described in Ref. \cite{Sca13b}. 
Equivalently to the monopole collective energy that depends of the quadrupole deformation, the onset of hexadecapole and higher order deformations create a coupling between  quadrupole and  higher order modes that modifies the collective energy.
A complete understanding of these effects might eventually be obtained using schematic models like fluid-dynamical models \cite{Nis85} or adiabatic cranking 
\cite{Abg80} but, due to the number of degrees of freedoms, this would lead to tedious formal development and discussions
that are out of the scope of the present article. 

%

\section{Nuclei with triaxial deformation} 

Among the nuclei considered in the present work, 54 are found to present triaxial deformation (see Fig. \ref{fig:map_def}).
To describe nuclei without symmetry axis, we follow  standard notations. In addition to the  parameter $\delta$ given by Eq. 
(\ref{eq:delta}), the angle $\gamma$ is introduced using the expression 
in terms of the static quadrupole moments:
\begin{align}
\cos(\gamma) &=  \frac{3  \langle Q_{20}\rangle }{4 A \langle r^2\rangle}, ~~~~
\sin(\gamma) =  \frac{3 \sqrt{3} \langle Q_{22}\rangle }{4 A \langle r^2\rangle}. \label{eq:gamma}
\end{align}  

\begin{figure}[!ht] 
	\centering\includegraphics[width=0.6\linewidth]{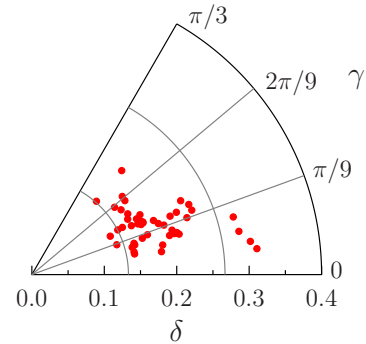}  
		\caption{(Color online) Position of the triaxial nuclei considered in the present work in 
		the $(\delta, \gamma)$ plane. Each point corresponds 
		to one nucleus.} 
	\label{fig:betagamma} 
\end{figure}

As an illustration of the diversity in the shapes considered here, we show in Fig. \ref{fig:betagamma}, the location of
triaxial nuclei in the $(\delta, \gamma)$ plane \cite{Boh97,Rin80}. 

\begin{figure*}[!ht] 
	\centering\includegraphics[width=0.8\linewidth]{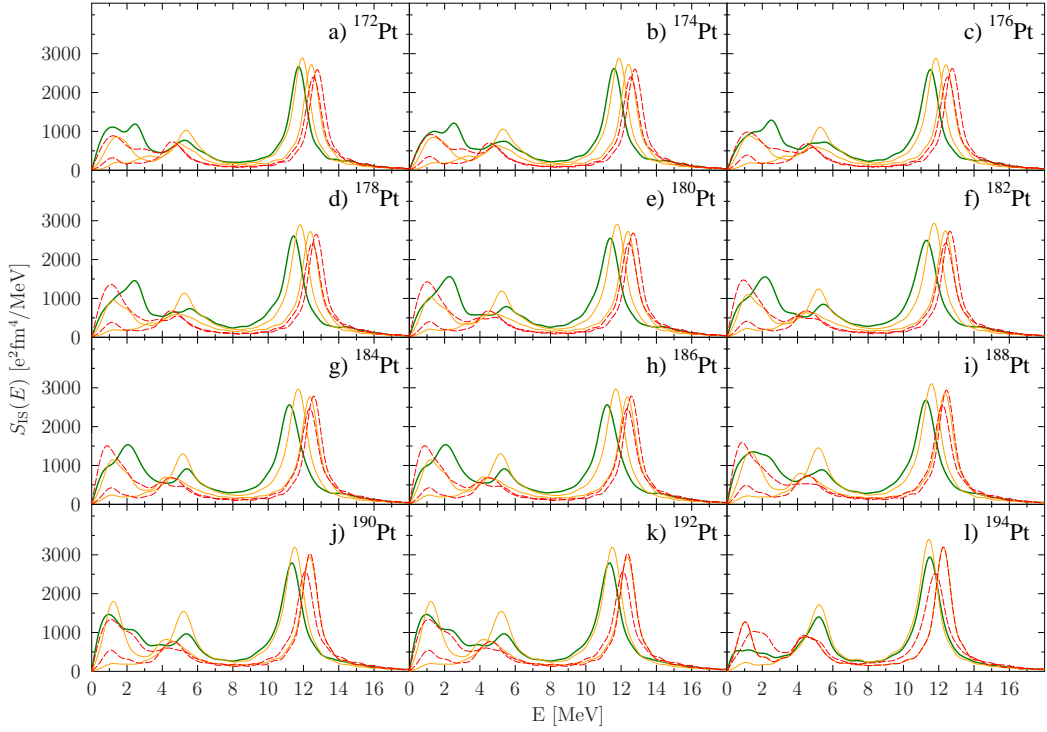}  
		\caption{(Color online) IS-GQR strength distributions obtained for the different $K$ projection 
		for the triaxial nuclei in the isotopic Pt chain. The $|K|=0$ response is shown by the thick green solid line, the two $|K|=1$
		and the two $|K|=2$ projections are shown respectively by the thin orange lines and red dashed lines. Note that a smoothing width 
		of 1 MeV has been used here to better see the peaks separation.} 
	\label{fig:ISPt} 
\end{figure*}
\begin{figure*}[!ht] 
	\centering\includegraphics[width=0.8\linewidth]{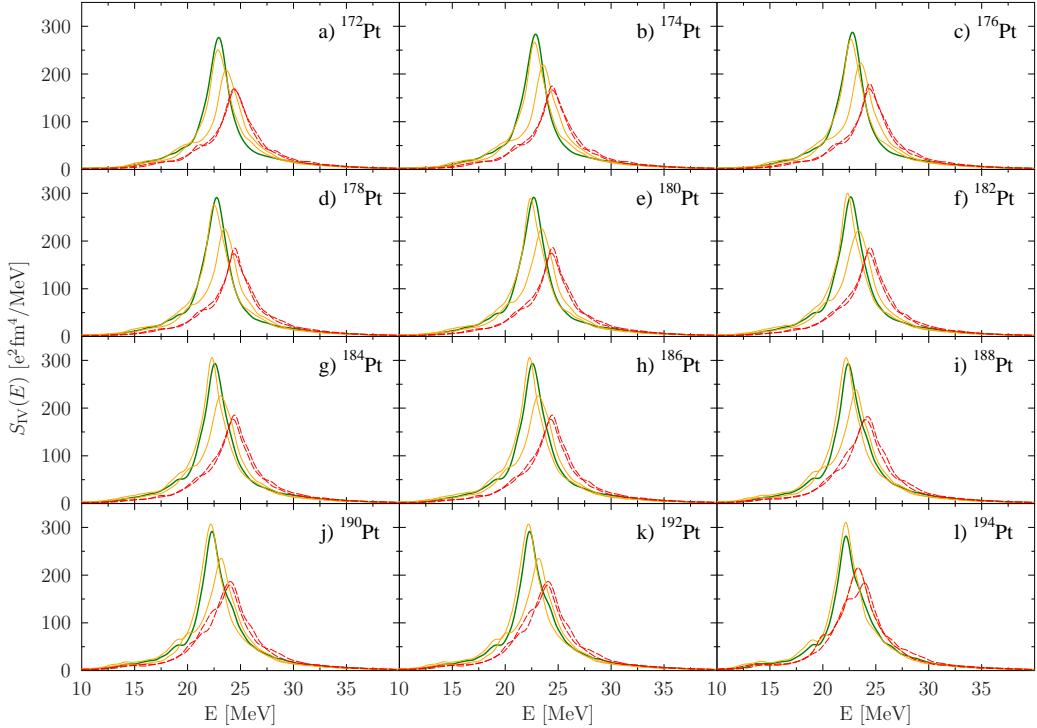}  
		\caption{(Color online) Same as figure \ref{fig:ISPt} for the IV-GQR.} 
	\label{fig:IVPt} 
\end{figure*}

Due to the absence of any symmetry axis, an extra splitting of the two $|K|=1$ and $|K|=2$ components is expected.  
Such splitting are illustrated in Figs. \ref{fig:ISPt} and \ref{fig:IVPt} for triaxial nuclei found in the Pt isotopic chain.  A clear splitting 
is seen in the $|K|=1$ channel and to a lesser extent in the $|K|=2$ case. We see in these figures that the extra effect of the absence of axial symmetry is rather weak compared to the net effect of deformation.  For instance,  the total width and the interplay between energy splitting and physical widths in different $|K|$ channels can be studied in a similar way as in axially deformed nuclei (see black points in Fig. \ref{fig:widthtot} and Fig. \ref{fig:widthIV}).The total width  in the IS-GQR completely follows the general trend observed for the case of axial deformation.

\begin{figure}[!ht] 
	\centering\includegraphics[width=0.9\linewidth]{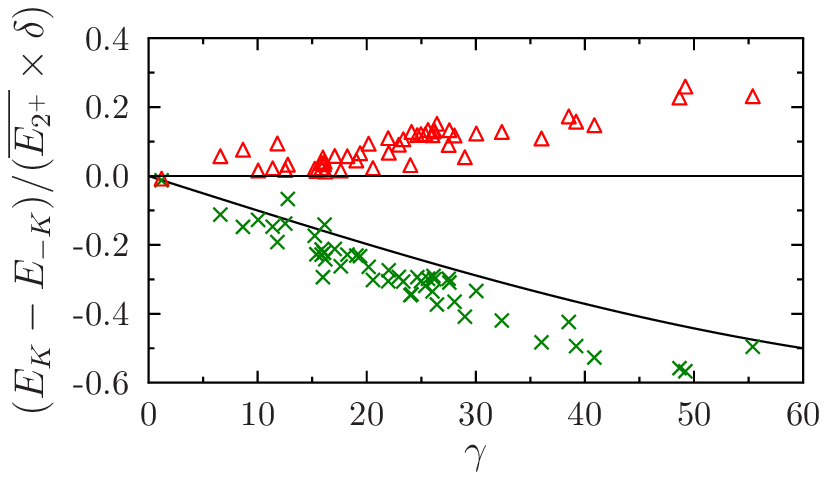}  
		\caption{(Color online) Energy difference between the two $|K|=1$ (green crosses)
		and the two $|K|=2$ (red open triangles) main collective peaks in triaxial nuclei as a function of 
		the deformation angle $\gamma$. The two lines correspond to the prediction of the adiabatic model 
		when only geometric effects in the diagonal part of the mass matrix are accounted for. For the $|K|=2$ 
		case (horizontal line) no splitting is predicted while for the $|K|=1$, Eq. (\ref{eq:splitk1}) is used where $E_{\rm sph}$
		is replaced by $\overline{E_{2+}}$.}
	\label{fig:splittriax} 
\end{figure}

The properties of the total strength distribution are largely dominated by the $\delta$ deformation parameters, nevertheless 
a weak but non-zero effects of $\gamma$ is observed especially in the $|K|=1$ case.  
In Figs. \ref{fig:splittriax} the extra splitting due to triaxiality is studied by showing the difference $\Delta E_{|K|} = (E_{K} - E_{-K})$
for $K=1$ and 2 as a function of the $\gamma$ parameter.  
In next section, the adiabatic method is used to give simple explanation 
to this extra splitting.

\begin{figure}[!ht] 
	\centering\includegraphics[width=0.9\linewidth]{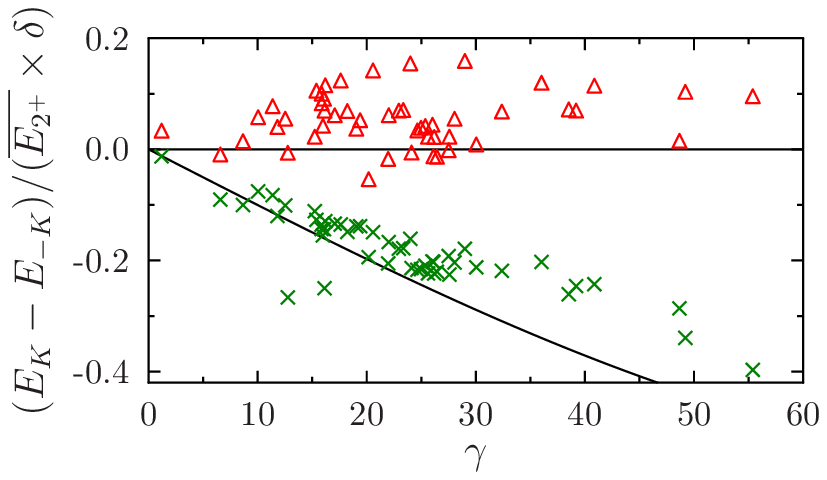}  
		\caption{(Color online) Same as Fig. \ref{fig:splittriax} for the IV-GQR. For the $|K|=2$ 
		case (horizontal line) no splitting is predicted while for the $|K|=1$, Eq. (\ref{eq:splitk1}) is used.} 
	\label{fig:splittriax_IV} 
\end{figure}

\subsection{Adiabatic approach to energy splitting in triaxial nuclei}

In the present work, we have followed the strategy of Ref. \cite{Abg80} that we found very useful to get 
better insight both in the geometrical properties of the different components of the IS-GQR and to understand the 
possible effect of  coupling with the GMR.  The aim of the present section is to provide a short summary of important 
aspects that were used to analyze the response obtained in the present article. 
Assuming that the IS-GQR and IS-GMR can be isolated from
other internal degrees of freedom and that anharmonic effects can be neglected, the hamiltonian describing the
collective motion can be generically written as:
\begin{align}
H = \sum_{i, j = 1}^6 \Big\{ -\frac{\hbar^2}{2M_{ij} } \frac{\partial^2}{\partial \alpha_i \partial \alpha_j} + \frac{1}{2} K_{ij} (\alpha_i - \overline{\alpha_i})(\alpha_j - \overline{\alpha_j}) \Big\}, \nonumber
\end{align} 
where $(\alpha_i, \dot \alpha_i)$ are the conjugated collective variables. In the following we will use $\eta$ for the 
collective variables associated to the monopole vibration, i.e. to the excitation operator $d_0=r^2$. $\{ \alpha_i \}_{i=-2,2}$ will 
be used for the quadrupole collective variables with the convention (see appendix \ref{app:excitation}):
\begin{align}
\alpha_0 &\leftrightarrow  d_{z^2}^{K=0},  \hspace*{0.3cm} 
\alpha_{-1}  \leftrightarrow   d_{xz}^{|K|=1}, \hspace*{0.3cm}  \alpha_{+1}   \leftrightarrow  d_{yz}^{|K|=1} \nonumber \\
\alpha_{-2}   &\leftrightarrow d_{xy}^{|K|=2}, \hspace*{0.3cm}   \alpha_{+2}   \leftrightarrow 
d_{x^2-y^2}^{|K|=2} .
\end{align}

The goal is not to redo the derivation performed in Ref. \cite{Abg80} but to summarize here important effects 
and eventually add new information. In the case of spherical nuclei, the off-diagonal matrix elements of the mass and of the restoring forces 
parameters cancel out and the collective Hamiltonian reduces to a set of independent oscillators. When deformation 
appears both diagonal part and off-diagonal matrix elements are modified. The adiabatic approach used  in  Ref. \cite{Abg80}
provides some indications of these modifications. An important finding is that the diagonal mass parameters can be related to the energy 
weighted sum rule discussed in appendix  \ref{sec:sumrule}. For off diagonal mass components, 
a generalized sum-rule can be used that is associated to couples of excitation operators. Using compact notations, we have:
\begin{align}
M_{ij} &= \frac{m}{4  } \langle \nabla d_i \nabla d_j \rangle= \frac{m^2}{2 \hbar^2} m^{d_i d_j}_1 . \label{eq:moff}
\end{align}
Note that in appendix \ref{app:excitation}, we focus on diagonal energy weighted sum rules and simply used the notation $m^{d_i d_i}_1 = m^{d_i}_1$.

\subsubsection{Geometric effects on the mass}

For the diagonal part of the mass, using expressions derived in appendix \ref{app:excitation}, we immediately obtain for the IS-GQR:
\begin{align}
\left\{
\begin{array}{l}
  M_{\eta} = m A \langle r^2 \rangle   \\
  M_{0} = m (2A \langle r^2 \rangle + \langle Q_{20} \rangle)    \\
  M_{-1} =   \frac{1}{2} m\left( 4 A \langle r^2 \rangle +  \langle Q_{20} \rangle + 3   \langle Q_{22} \rangle \right) \\
  M_{+1} =   \frac{1}{2} m \left( 4 A \langle r^2 \rangle +  \langle Q_{20} \rangle - 3   \langle Q_{22} \rangle \right) \\
  M_{-2} = m (2A \langle r^2 \rangle - \langle Q_{20} \rangle)   \\
  M_{+2} =  m (2A \langle r^2 \rangle - \langle Q_{20} \rangle)
\end{array}
\right. .\label{eq:massdiag}
\end{align}   
Note that, similar strategy can be used also in the IV-GQR but leads to slightly more complex expressions.

\begin{figure}[!ht] 
	\centering\includegraphics[width=0.9\linewidth]{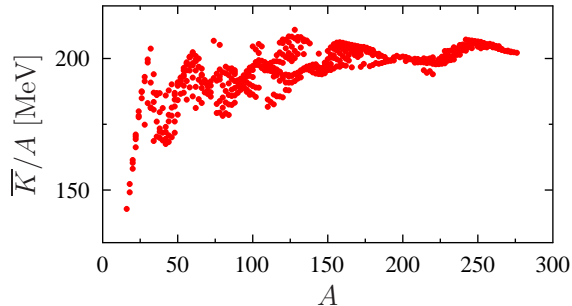}  
		\caption{(Color online) restoring force of the IS-GQR computed with the formula (\ref{eq:Polarisabilty}) as a function of the mass. } 
	\label{fig:polarisability} 
\end{figure}

General arguments regarding the restoring force turn out to be more complex.   Nevertheless, in spherical nuclei, 
knowing the collective energy and the mass, an order of magnitude of the restoring force in finite nuclei can simply be obtained 
using:
\begin{align}
E_{i} = \hbar \sqrt{\frac{K_i}{M_i}} .\label{eq:esimple}
\end{align} 
In spherical nuclei, the different $|K|$ components have same energies and collective masses. Accordingly, we also 
have $K_0 = K_1 = K_2$. To get an order of magnitude of the average restoring force for all nuclei, including deformed ones, 
we can define the quantity $\overline{K}$, through 
\begin{align}
\frac{\overline{K}}{A}=\frac1{5A} \sum_i K_i = \frac1{5A} \sum_{K=-2}^{+2}  \frac{ E^2_{K} M_K }{\hbar^2}
\label{eq:Polarisabilty}
\end{align} 
Using the fitted collective energy for each $|K|$ and the diagonal masses given above, an estimate of $\overline{K}$ can be made.
This quantity is shown as a function of the mass for spherical and deformed nuclei on Fig. \ref{fig:polarisability}. 
While important fluctuations exists in the low mass region, for mass $A>100$, the quantity $\frac{\overline{K}}{A}$ becomes 
almost constant with a value around 200 MeV. 
Note that, due to the collective energy dependence of the Skyrme functional \cite{Sca13a}, 
the restoring force strongly depends on the specific interaction used in the mean-field channel. 

The deformation affects the GQR response in several ways. The most evident effect comes from 
the change in the diagonal part of the mass. Their dependence in terms of deformation parameters 
and the influence on the collective energies 
can directly be understood using the expressions (\ref{eq:massdiag})  in Eq. (\ref{eq:esimple}) together with the  
definition of $\delta$ eq. (\ref{eq:delta}) and $\gamma$ eq. (\ref{eq:gamma})), we deduce: 
\begin{align}
\left\{
\begin{array}{l}
~~E_{0} = E_{\rm sph} \left( 1 + \frac{2}{3} \delta \cos(\gamma) \right)^{-1/2}  \\
E_{-1} = E_{\rm sph} \left( 1 + \frac{1}{3} \delta \cos(\gamma) + \frac{1}{\sqrt3} \delta \sin(\gamma) \right)^{-1/2}  \\
E_{+1} = E_{\rm sph} \left( 1 + \frac{1}{3} \delta \cos(\gamma) - \frac{1}{\sqrt3} \delta \sin(\gamma) \right)^{-1/2}  \\
E_{-2} = E_{\rm sph} \left( 1 - \frac{2}{3} \delta \cos(\gamma)  \right)^{-1/2}  \\
E_{+2} = E_{\rm sph} \left( 1 - \frac{2}{3} \delta \cos(\gamma)  \right)^{-1/2}  
\end{array}
\right. ,
\label{eq:Egammadelta}
\end{align}   
where $E_{\rm sph} = \sqrt{K_0 /(2Am \langle r^2 \rangle) }$.

For small deformations, these expressions can then be expanded in $\delta$.
Note that at first order in $\delta$, for axially 
symmetric systems ($\gamma=0$), we recover expressions (\ref{eq:0delta}-\ref{eq:2delta}). 
It is interesting to mention that the above formulas predict a splitting of the $|K|=1$ components 
due to the change of the collective masses, while no splitting of the $|K|=2$ components is anticipated. 
More precisely, defining by $\Delta E_1 = E_{K=1} - E_{K=-1}$, we obtain:
\begin{align}
\Delta E_1 \simeq  - E_{\rm sph} \frac{\delta}{\sqrt{3}} \sin{\gamma} . \label{eq:splitk1}
\end{align} 
This expression turns out to explain rather well the trends observed in Fig. \ref{fig:splittriax}.  
Here, we do not treat the IV-GQR that will be slightly more involved 
due to the more complex expression of the sum-rules given in appendix \ref{app:excitation}. Nevertheless, 
Fig.  \ref{fig:splittriax_IV} shows that the $|K|=1$ splitting in the IV channel 
 is also compatible with the dependence (\ref{eq:splitk1}).
We also see that some residual deviations occur especially at large $\gamma$. Below, we investigate if the 
inclusion of coupling between different vibrational modes can further improve the description.

\subsubsection{Coupling between different collective vibrational modes}

\begin{figure}[!ht] 
	\centering\includegraphics[width=0.9\linewidth]{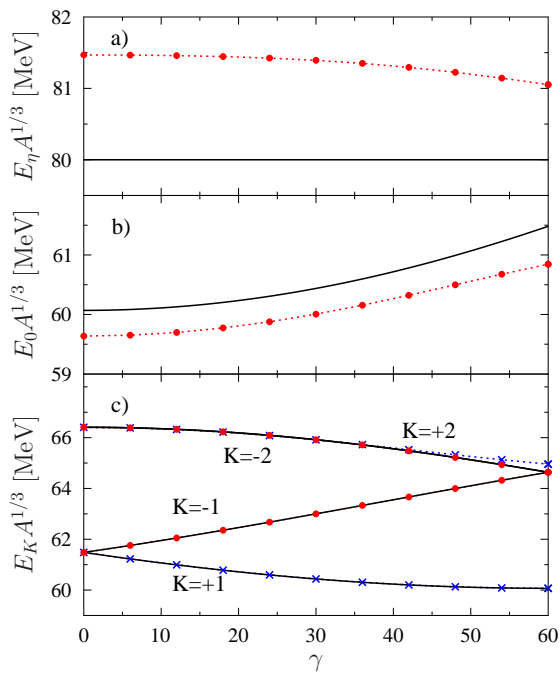}  
	\caption{(Color online) Evolution of the energies of different modes: a) monopole , b) $K=0$ c) $K= \pm 1$ and $K= \pm 2$ as a function of the $\gamma$ parameter. In all cases, the calculation 
	neglecting the couplings (black solid line) are compared to the case with coupling (red filled circles). 
	In panel c), the two components without coupling for $|K|=1$ and $2$ are shown with solid lines. When 
	coupling is included, the negative projection is displayed by red filled circles while the positives components 
	are shown by blue crosses.  Note that the 
	$|K|=1$ components are not affected by the coupling.} 
	\label{fig:coupling} 
\end{figure}

Besides the geometric effects on the diagonal masses. Deformation will also lead to a coupling between different collective 
modes associated to the off-diagonal components $(K_{ij};M_{ij})$. Simple analysis in the adiabatic approach show that some of the restoring 
force matrix elements cancel out while others are very small compared to the diagonal terms (see table I of Ref. \cite{Abg80}). Here we will 
assume that the $\{ K_{ij} \}_{i\neq j}$ components can be simply be neglected. 
According to this hypothesis and to the symmetry imposed in the {\rm EV8} program, 
only three off-diagonal elements of the mass matrix are different from zero. Starting from Eq. (\ref{eq:moff}), one obtain
\begin{align}
\left\{
\begin{array}{l}
  ~~M_{\eta 0} = m  \langle Q_{20} \rangle  \\
  M_{\eta +2} = m \langle Q_{22} \rangle \\
  M_{0 +2} =  -m  \langle Q_{22} \rangle  
\end{array}
\right. .\label{eq:mass}
\end{align}   
Then, adding the coupling between vibrational modes 
will not affect the  $|K|=1$ components but will modify the $K=0$ and $K=\pm 2$ energies. In particular since, only $K=+2$
is coupled with the IS-GMR and with the $K=0$ components, the degenerescence of the two $|K|=2$ modes is removed.  To quantify the 
effect of the coupling, using different parameters introduced above, the collective hamiltonian can be directly diagonalized in 
the collective space. In Fig. \ref{fig:coupling}, an illustration of the coupling effect as a function of $\gamma$ is shown. This figure has been 
performed for a mass $A=180$, that corresponds to the mass region of the Pt isotopes shown in Fig. \ref{fig:ISPt} and for
a deformation $\delta =0.15$ that is the average value obtained for triaxial nuclei (see Fig. \ref{fig:betagamma}).  In addition, a simple formula $E_\eta = 80 A^{-1/3} [MeV]$  has been assumed for the IS-GMR before coupling. 
The main effect of coupling is to shift down (resp. up) the $K=0$ IS-GQR (resp. the IS-GMR) collective energy. Conjointly, we anticipate 
that a peak appears in the IS-GQR response with energy corresponding to the shifted IS-GMR energy. For the Pt isotopes shown in 
Fig. \ref{fig:ISPt}, this would correspond to a peak around $E\simeq 14-15$ MeV. A careful analysis of the green solid curves 
in Fig  \ref{fig:ISPt}, shows that a small component indeed appears in this energy range. 

In addition to these shifts, at large deformation, a splitting of the two $|K|=2$ components is observed. This is due to the 
fact that only the $K= +2$ collective peak is coupled to other vibrational mode. However, this splitting is rather weak and is not enough to explain the  $|K|=2$ splitting observed in Fig. \ref{fig:splittriax}.

\section{Conclusion}
In the present work, the IS-GQR and IV-GQR response of deformed nuclei is systematically investigated 
over the whole nuclear chart using the time-dependent EDF. The changes in the GQR response functions 
due to deformation are carefully analyzed.  When nuclei with an axis of symmetry are considered a splitting of the 
GQR are observed. This splitting is accompanied by a change in the relative weight of the different $|K|$ components 
of the response. It is shown that this splitting can be rather well understood using the fluid dynamical model of 
Ref. \cite{Nis85}.  However, non-trivial effects beyond this simple modelization are uncovered.  It is shown that the possible 
occurrence of high order multipole deformations induces significant fluctuations of the mean-collective energy with respect 
to the fluid dynamical approach. In addition, contrary to spherical nuclei, as the deformation increases, the IS-GQR in medium- and 
heavy-nuclei acquires a non negligible physical spreading width for all $|K|$ components that significantly contributes to the overall fragmentation of the strength. A detailed study is also made for triaxial nuclei. 
All new effects observed for axially symmetric nuclei are also seen in the absence of symmetry axis, 
i.e. splitting of the GQR response and damping. In addition, an extra splitting between the negative and positive components 
that is studied quantitatively is observed. A detailed analysis of new aspects occurring in triaxial nuclei is made. In particular, the 
interplay between geometric aspects and coupling between different degrees of freedom is made. 

The present work, besides the physical insight in the GQR process, demonstrates that the time-dependent EDF 
can provide an interesting tool for systematic studies where other techniques become rather tedious.  

Last, it is interesting to mention that the onset of deformation increases significantly the complexity in the fragmentation 
and damping of the collective modes with the appearance of fluctuations at different scales.  An interesting continuation of the 
present work could be to apply the wavelet analysis proposed in Refs. \cite{Lac99,Lac00}  and further improved in Refs. \cite{Sch04, Sch09} in order 
to quantitatively study the different energy scales induced by deformation.

Note that all the strength functions of the deformed nuclei considered here are available in the supplement material \cite{Sup13}.

\section*{Acknowledgment}  
We would like to thank T. Nakatsukasa and K. Yoshida for their help and discussions in the comparison with QRPA.
 
 \appendix
 
 \section{GQR excitation operators}
\label{app:excitation} 
 
The GQR response within QRPA is generally studied using excitation operators written in terms 
of the spherical harmonics:   
\begin{align}
Q_{2K}^{IS} &= e \sum_{i=1}^{A} r^2_i Y_{2K}(\Omega_i) \\
Q_{2K}^{IV} &= \frac{eN}{A} \sum_{i=1}^{Z} r^2_i Y_{2K}(\Omega_i) - \frac{eZ}{A} \sum_{i=1}^{N} r^2_i Y_{2K}(\Omega_i),
\end{align}
with $K= 0$, $\pm 1$ and $\pm 2$.
For spherical nuclei, the response function obtained for different $K$ are identical, for deformed nuclei this is not true 
anymore. Here, for the sake of simplicity in the boost applied prior to the evolution, 
we use similar expressions except that the spherical harmonics are replaced by the real spherical harmonics: 
\begin{align}
d_{z^2}^{K=0} &= Y_{20} =  \frac14 \sqrt{\frac{5}{\pi}} \frac{-x^2-y^2+2z^2}{r^2} \\
d_{xz}^{|K|=1} &=  \sqrt{\frac{1}{2}} \left(Y_{2-1} - Y_{21}\right) = \frac12 \sqrt{\frac{15}{\pi}} \frac{xz}{r^2} \\
d_{yz}^{|K|=1} &=  i\sqrt{\frac{1}{2}} \left(Y_{2-1} + Y_{21}\right) = \frac12 \sqrt{\frac{15}{\pi}} \frac{yz}{r^2}  \\
d_{xy}^{|K|=2} &= i \sqrt{\frac{1}{2}} \left(Y_{2-2} - Y_{22}\right) =  \frac12 \sqrt{\frac{15}{\pi}} \frac{xy}{r^2} \\
d_{x^2-y^2}^{|K|=2} &= \sqrt{\frac{1}{2}} \left(Y_{2-2} + Y_{22}\right) =  \frac14\sqrt{\frac{15}{\pi}} \frac{x^2-y^2}{r^2} ,
\end{align}
leading to two sets of 5 operators, denoted respectively by $Q_{z^2}^{IS/IV}$, $Q_{xz}^{IS/IV}$, 
$Q_{yz}^{IS/IV}$, $Q_{xy}^{IS/IV}$ and $Q_{x^2 - z^2}^{IS/IV}$. The connection between the strength functions obtained with the standard spherical 
harmonics and the real spherical harmonics are rather straightforward. Denoting generically $S$ the strength distribution and using natural notations, we have:
\begin{align}
S_{z^2} &= S_{20} \\
S_{xz} + S_{yz} &= S_{21} +S_{2-1}  \\
S_{xy} + S_{x^2-y^2} &= S_{22} +S_{2-2} .
\end{align}  

If in addition, the system is axially symmetric, we finally simply have:
\begin{align}
S_{xz} = S_{yz} &= S_{21} = S_{2-1} \\
S_{xy} = S_{x^2-y^2} &= S_{22} = S_{2-2} .
\end{align}

\subsection{Sum rules}
\label{sec:sumrule}

For an isoscalar operator $F$ the energy weighted sum-rule, denoted by $m_1(F)$
 can directly be deduced using the general formula \cite{Boh97}
\begin{align}
m_1^F &= \sum_{a} (E_{a}-E_0) |\langle a| F | 0 \rangle |^2 \nonumber \\
 &= \frac12 \langle 0 | [F,[H,F]] | 0 \rangle \nonumber \\
&= \frac{\hbar^2}{2m} \langle 0 | \sum_k |{\vec \nabla}_k F({\bf r}_k)|^2 | 0 \rangle .
\end{align}
For the considered operator, we obtain:
\begin{align}
m^{z^2}_1 
&= e^2 \frac{5\hbar^2 }{8 \pi m} \left( 2 A \langle r^2 \rangle +  \langle Q_{20} \rangle \right)  \\
m^{xz}_1 &=  e^2\frac{5 \hbar^2}{16 \pi m}  \left( 4 A \langle r^2 \rangle +  \langle Q_{20} \rangle + 3   \langle Q_{22} \rangle \right) \\
m^{yz}_1 &=  e^2 \frac{5 \hbar^2}{16 \pi m} \left( 4 A \langle r^2 \rangle +  \langle Q_{20} \rangle - 3   \langle Q_{22} \rangle \right)  \\
m^{xy}_1 &= e^2 \frac{ 5  \hbar^2}{8 \pi m}  \left( 2 A \langle r^2 \rangle -  \langle Q_{20} \rangle  \right)
 \\
m_1^{x^2 - y^2} &= e^2 \frac{5  \hbar^2}{8 \pi m}   \left( 2 A \langle r^2 \rangle -  \langle Q_{20} \rangle  \right) .
\end{align}
Note that for the case of spherical symmetric nuclei with $ \langle Q_{20} \rangle =  \langle Q_{22} \rangle=0$, 
these sum-rules are all equals. For axially symmetric nuclei with  $ \langle Q_{22} \rangle=0$ we have the additional properties 
$m^{xy}_1 = m^{xz}_1$ {and $m^{xz}_1 = m^{x^2-y^2}_1$}. 

For the IV excitation, we 
can make use of the general formula valid when the effective interaction has a momentum dependent part \cite{Ter06}
\begin{align}
S(F) &= \frac{\hbar^2}{2m} \left[ \int |\nabla F|^2 \rho_n({\bf r}) d{\bf r}
+ \int |\nabla F|^2 \rho_p({\bf r}) d{\bf r} \right]   
 \nonumber \\
&+ C  \int |\nabla F|^2 \rho_n({\bf r}) \rho_p({\bf r}) d{\bf r} ,
\end{align}
where 
\begin{align}
C &= \frac14( t_1 ( 1 + \frac12 x_1 ) + t_2 ( 1 + \frac12 x_2 ) ). 
\end{align}
\begin{widetext}

Leading to the different isovector sum rules: 
\begin{align}
m^{z^2}_1 &=  e^2 \frac{5 \hbar^2}{8 \pi m} \left( \frac{N^2}{A^2} \left(2 Z \langle r^2 \rangle_p +  \langle Q_{20} \rangle_p \right) + \frac{Z^2}{A^2}  \left(2 N \langle r^2 \rangle_n +  \langle Q_{20} \rangle_n \right) \right) \nonumber \\
&+ C \frac{5}{16\pi} \int  (16z^2+4y^2+4x^2 )  \rho_n({\bf r}) \rho_p({\bf r}) d{\bf r}  \\
m^{xz}_1&= e^2 \frac{5 \hbar^2}{16 \pi m} \left( \frac{N^2}{A^2} \left(4 Z \langle r^2 \rangle_p +  \langle Q_{20} \rangle_p + 3  \langle Q_{22} \rangle_p \right) + \frac{Z^2}{A^2}  \left(4 N \langle r^2 \rangle_n +  \langle Q_{20} \rangle_n + 3  \langle Q_{22} \rangle_n \right) \right) \nonumber \\
&+ C \frac{15}{4\pi} \int (z^2+x^2 ) \rho_n({\bf r}) \rho_p({\bf r}) d{\bf r} \\
 m_1^{yz}&= e^2 \frac{5 \hbar^2}{16 \pi m} \left( \frac{N^2}{A^2} \left(4 Z \langle r^2 \rangle_p +  \langle Q_{20} \rangle_p - 3  \langle Q_{22} \rangle_p \right) + \frac{Z^2}{A^2}  \left(4 N \langle r^2 \rangle_n +  \langle Q_{20} \rangle_n - 3  \langle Q_{22} \rangle_n \right) \right) \nonumber \\
&+ C \frac{15}{4\pi} \int (z^2+y^2 ) \rho_n({\bf r}) \rho_p({\bf r}) d{\bf r}  \\
m^{xy}_1  &=e^2 \frac{5 \hbar^2}{8 \pi m}  \left( \frac{N^2}{A^2} \left(2 Z \langle r^2 \rangle_p -  \langle Q_{20} \rangle_p \right) + \frac{Z^2}{A^2}  \left(2 N \langle r^2 \rangle_n -  \langle Q_{20} \rangle_n \right) \right) \nonumber \\
&+ C \frac{15}{4\pi} \int (y^2+x^2 ) \rho_n({\bf r}) \rho_p({\bf r}) d{\bf r}  \\
 m^{x^2-y^2}_1&=  e^2 \frac{5 \hbar^2}{8 \pi m} \left( \frac{N^2}{A^2} \left(2 Z  \langle r^2 \rangle_p -  \langle Q_{20} \rangle_p \right) + \frac{Z^2}{A^2}  \left(2 N \langle r^2 \rangle_n -  \langle Q_{20} \rangle_n \right) \right) \nonumber \\
&+ C \frac{15}{4\pi} \int (y^2+x^2 ) \rho_n({\bf r}) \rho_p({\bf r}) d{\bf r}.
\end{align}
\end{widetext}

 \section{Removal of spurious rotation in $|K|=1$ channel}
 \label{sec:rotation}
 \begin{figure}[!ht] 
	\centering\includegraphics[width=\linewidth]{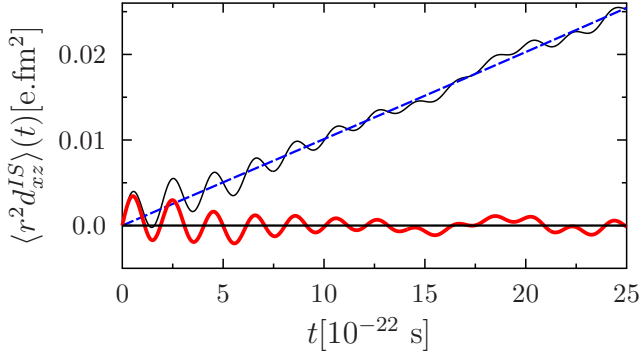}  
	\caption{ (Color online) Evolution of the quantity $\langle r^2 d_{xz}^{|K|=1} \rangle $ in $^{24}$Mg after a $|K|=1$ boost (black thin solid line). The 
	blue dashed line corresponds to a fit with a third order polynomial $P(t)$ while the red thick line corresponds to 
	$[\langle r^2 d_{xz}^{|K|=1} \rangle  - P(t) ]$. The horizontal line at $y=0$ is added to guide the eyes.} 
	\label{fig:meanMg24} 
\end{figure}
\begin{figure}[!ht] 
	\centering\includegraphics[width=\linewidth]{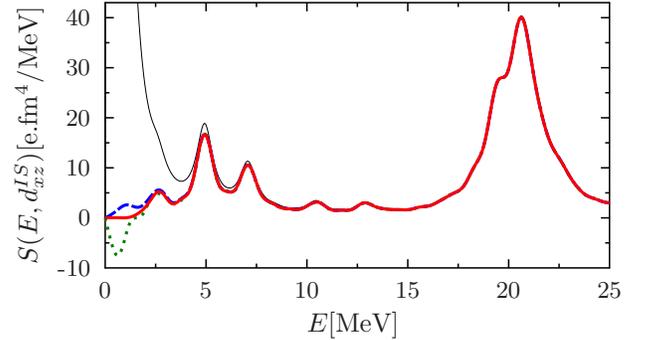}  
	\caption{ (Color online) Response function associated to a $|K|=1$ boost with $F=r^2 d_{xz}^{|K|=1}$ excitation operator.  
	The response in black thin solid line is obtained including the rotation. To properly get the low energy part, it was necessary to perform the evolution up to a large finite time $T=900\times 10^{-22} s$. Illustration of the strength distribution obtained using a first order (green dotted line), second order (blue long dashed line) and third order polynomial (red thick solid line) 
	fit that is removed from the evolving quantity. In all cases, the final time used 
	is much shorter ($T = 50\times 10^{-22} s$). } 
	\label{fig:Mg_24_correction} 
\end{figure}

When a $|K|=1$ boost operator associated for instance to the operator  $F=r^2 d_{xz}^{|K|=1}$ is used in deformed nuclei, a time-dependent rotation of the whole nucleus is observed as a function of time in the intrinsic frame. This situation is illustrated 
in Fig. \ref{fig:meanMg24} where the mean value of $F$ is displayed as a function of time. The rotation is associated to a slow 
increase of $\langle F \rangle$.  
This rotation leads to a spurious very low frequency component shown in Fig. \ref{fig:Mg_24_correction} (black thin solid line).
 
To get rid of the spurious rotation, we propose a rather simple method consisting in fitting the 
evolving quantity by a polynomial function,
\begin{align}
P(t) = a_1 t + a_2 t^2 +a_3 t^3 + ... 
\end{align}
and then remove this function from the considered quantity before performing the Fourier transform. 
In practice, we found that a third order polynomial is enough to completely remove the rotation contribution. 
An illustration of the fit is given in Fig.  \ref{fig:meanMg24}. The corresponding effect on the strength distribution 
is shown in Fig. \ref{fig:Mg_24_correction}. We see that the methods leaves the high energy spectra unchanged while removing 
completely the unphysical low energy component.

\section{A formula connecting different widths}

\label{sec:width}

For simplicity, let us assume that each peak is properly described by a gaussian with a single peak of 
height $H_K$. We assume:
\begin{align}
S_K(E) &= \frac{H_K}{\sqrt{2\pi} \sigma_K} e^{-\frac{(E-E_K)^2}{2\sigma^2_K}} .
\end{align}
With this definition, we have
\begin{align}
\int_{-\infty}^{+\infty} S_K(E) dE &= H_K  \\
\int_{-\infty}^{+\infty} E S_K(E) dE &= H_K E_K  \\
\int_{-\infty}^{+\infty} E^2 S_K(E) dE &= H_K ( \sigma^2_K + E^2_K ) .
\end{align} 
We consider the sum of the different $K$ quantities as:
\begin{align}
S(E) = \sum_{K=-2} ^{+2} S_K (E) .
\end{align}
From this we deduce:
\begin{align}
\int_{-\infty}^{+\infty} S(E) dE &= \sum_K H_K   \\
\int_{-\infty}^{+\infty} E S(E) dE &= \sum_K H_K E_K  \\
\int_{-\infty}^{+\infty} E^2 S(E) dE &= \sum_K H_K (\sigma^2_K + E^2_K) .
\end{align}
We can define the average energy and width:
\begin{align}
H_{\rm tot} &= \sum_K H_K  \\
{E}_{\rm tot} &= \left( \sum_K c_K E_K  \right)  \\
\sigma^2_{\rm tot} &= \frac{1}{H_{\rm tot}}  \int_{-\infty}^{+\infty} E^2 S(E) dE  \nonumber \\
&- \left( \frac{1}{H_{\rm tot}}  \int_{-\infty}^{+\infty} E S(E) dE \right)^2 \nonumber \\
&= \sum_K c_K \sigma^2_K + \sigma^2_E ,
\end{align}
where we have introduced the notation $c_K = H_K/H_{\rm tot}$ and where 
\begin{align}
\sigma^2_E &\equiv  \sum_K c_K E^2_K - \left(\sum_K c_K E_K \right)^2 .
\end{align}
We see here that there are two effects contributing to the total width: the weighted addition 
of the different individual width and the spreading of the peaks.

Eventually, we can assume that a set of Gaussian can be replaced by a single Gaussian written as
\begin{align}
S(E) &= \frac{H_{\rm tot}}{\sqrt{2\pi} \sigma_{\rm tot}} e^{-\frac{(E-E_{\rm tot})^2}{2\sigma^2_{\rm tot}}} .
\end{align}
The connection with the $\Gamma_K$ can be made by imposing that the Gaussian is centered at the same energy and 
has the same full width at half 
maximum as the Lorentzian. Then we have for each $\Gamma_K$:
\begin{align}
\Gamma_K = 2 \sigma_K \sqrt{2 \ln 2}, 
\end{align}
and 
\begin{align}
\Gamma_{\rm tot} &= 2 \sigma_{\rm tot} \sqrt{2 \ln 2}.
\end{align}
In terms of $\Gamma$, we deduce the relationship:
\begin{align}
\Gamma^2_{\rm tot} &= \sum_K c_K  \Gamma^2_K +  \Gamma^2_E, \label{eq:fiteqaa}
\end{align}
where $\Gamma^2_E \equiv 8  \ln 2 \sigma^2_E $.

\end{document}